\documentclass[aps,twocolumn,showpacs]{revtex4}

\usepackage{graphicx}
\usepackage{amsmath,amssymb}

\def \E {{\cal E}}

\begin{document}
\title{Storing and releasing light in a gas of moving atoms}
\author{G. Juzeli\={u}nas, M. Ma\v{s}alas}
\affiliation{Institute of Theoretical Physics and Astronomy,
Vilnius University, A. Go\v{s}tauto 12,
Vilnius 2600, Lithuania}
\author{M. Fleischhauer}
\affiliation{Fachbereich Physik, Universit\"at Kaiserslautern, D-67663\\
Kaiserslautern, Germany}
\date{\today}

\begin{abstract}
We propose a scheme of storing and releasing pulses or
cw beams of light in
a moving atomic medium illuminated by two
stationary and spatially separated control
lasers. The method is based on electromagnetically induced transparency
(EIT) but in contrast to previous schemes, storage and retrieval
of the probe pulse can be achieved at different locations and without
switching off the control laser.
\end{abstract}


\maketitle

Over the past couple of years there has been a great deal of interest in
storing and releasing light pulses in atomic gases exhibiting
electromagnetically induced transparency (EIT)
\cite{Fleisch:00,Hau:01,Lukin:01,Scully:01,Juz:02,Fleisch:02,Scully:02}. EIT
is a phenomenon in which a weak probe pulse travels slowly and almost
without dissipation in a resonant medium controlled by another
laser 
\cite{Harris:90,Harris:91,Arimondo:96,Harris:97,Scully:book,Bergman:98,
Hau:99}.
Following the proposal of Ref.\cite{Fleisch:00},
storage and release of a probe pulse
has been demonstrated \cite{Hau:01,Lukin:01,Hemmer:02} by
dynamically changing the intensity of the control laser.
In the read-out process it is possible to alter certain properties of the
stored probe pulse, such as its carrier frequency, propagation direction
or pulse shape \cite{Juz:02,Scully:02}.

Recently an application of the light-storage scheme
to generate continuous
beams of {\it atoms} in  nonclassical quantum states was proposed
\cite{Gong}. In
that scheme the probe field propagates co-parallel to a beam of atoms in
a spatially varying control field. This corresponds in the rest frame of the
atoms to an explicitly time dependent control laser and thus
allows for a complete and loss-free adiabatic transfer of probe-field
excitations to the matter wave \cite{Fleisch:02}.
We consider here another situation in which both the
control and the incoming probe beams are perpendicular to the atomic
motion. Such a scheme is better suited to eliminate effects of
Doppler broadening and thus to alleviate limitations to the allowed velocity
spread of the atomic beam.
Furthermore
in contrast to previous setups
\cite{Fleisch:00,Hau:01,Lukin:01,Scully:01,Juz:02,Fleisch:02,Scully:02}, 
storing and releasing of a probe beam can now be achieved with a pair of stationary control 
lasers, i.e. there is no need to
switch "off" and "on" a control laser at precise times. This is
advantageous if one does not know the exact time of arrival of the probe
pulse. Thus the present set-up could be used for example to store
entangled photons generated
by a continuous optical parametric oscillator below threshold and frequency selected by a 
cavity
\cite{Schori02,Lvovsky01}.
Finally storage and release of the probe field are spatially separated in 
the
present scheme.

Consider a stream of atoms moving along the $x$ axes as shown in fig.1 (bottom part).
The atoms are
characterized by two hyper-fine ground levels $g$ and $q$, as well as an
electronic excited level $e$ (top part of fig.1). 
The atoms in different internal states
are described by the field-operators $\Psi _{g}\equiv
\Psi _{g}\left( {\bf r},t\right) $,
$\Psi _{q}\equiv \Psi _{q}\left( {\bf r},t\right) $ and $\Psi
_{e}\equiv \Psi _{e}\left( {\bf r},t\right) $ obeying Bose-Einstein or
Fermi-Dirac commutation relations (depending on the type of atoms). The
matter fields interact with several light fields propagating along the
$z $-axis. A strong classical control laser centered at $x=x_{1}$
drives the transition $\left| e\right\rangle \rightarrow \left|
q\right\rangle $, whereas a weaker quantum probe field (also entering the
medium at $x=x_{1}$) is coupled with the transition $\left| g\right\rangle
\rightarrow \left| e\right\rangle $. Finally there is a second spatially
separated control laser (centered at $x=x_{2}$) that is used to release the
stored probe pulse.

%%%%%%%%%%%%%%%%%%%%%%%%%%%%%%%%%%%%%%%%%%%%%%%%%%%%%%%%%%%%%%%%%%%%%%%%%
%
%
\begin{figure}[ht]
\begin{center}
\includegraphics[width=5cm]{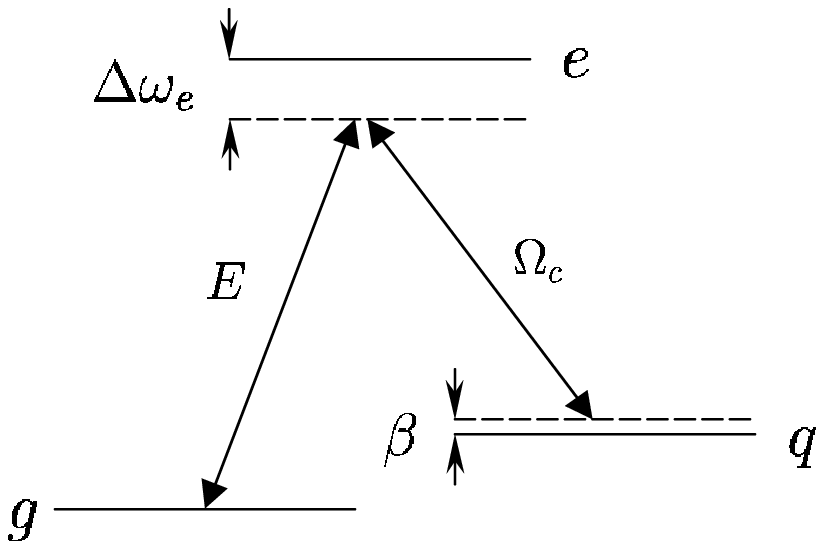}
\includegraphics[width=8.5cm]{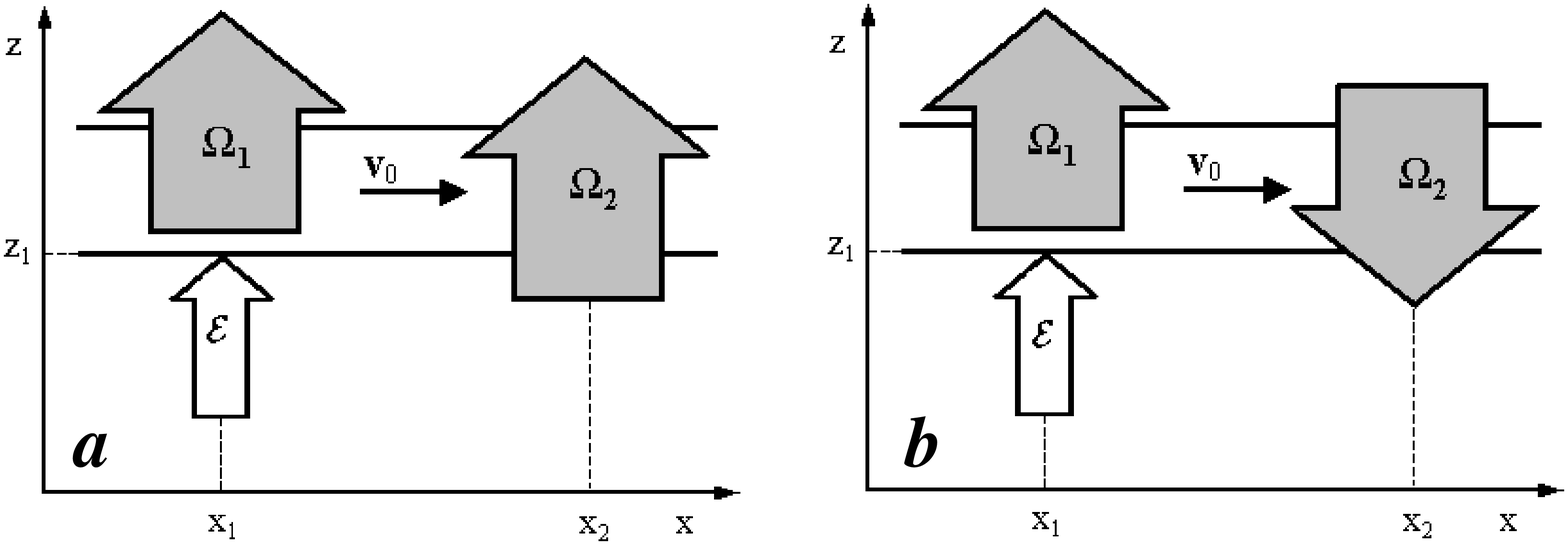}
\caption{{\it top:} Atomic level scheme. {\it bottom:}
A stream of atoms moving along the $x$ axes  with an
average speed ${\bf v}_{0}$. Two strong classical control lasers
characterized by the Rabi frequencies $\Omega _{1}$ and $\Omega _{2}$ are
centered at $x=x_{1}$ and $x=x_{2}$, respectively. The first control beam
propagates along the $z$-axis. The second beam propagates either parallel
(a) or anti-parallel (b)
to the  $z$ axis. A quantum probe field ($\E$) enters the medium at
$x=x_{1}$.
}
\label{setup}
\end{center}
\end{figure}
%
%%%%%%%%%%%%%%%%%%%%%%%%%%%%%%%%%%%%%%%%%%%%%%%%%%%%%%%%%%%%%%%%%%%%%%%%%%%

The first control laser has a frequency $\omega_{c} $,
a wave vector ${\bf k}_{c}=\widehat{{\bf z}}k_{c}$, and a Rabi
frequency
%
%
%\begin{equation}
$\Omega \left( {\bf r},t\right) =\Omega _{1}\left( x\right) e^{-i\left(
\omega _{c}t-k_{c}z\right) }, $
% \label{Omega}
% \end{equation}
%
%
where the $y$ dependence of the amplitude $\Omega _{1}\left( x\right) \equiv
\Omega _{1}$ will be kept implicit. The same applies to other parameters,
such as the mean atomic number density $n\equiv n\left( z\right) $.

The probe beam is described by the electric field operator:
%
%
%\begin{equation}
${\bf E}\left( {\bf r},t\right) =
\widehat{\bf e}\left(\hbar ck /2\varepsilon
_{0}\right) ^{1/2}\E\left( {\bf r},t\right) e^{-i\left( \omega t-kz\right)
}+H.c. $
% \label{Ansatz-E}
% \end{equation}
%
%
where $\E\left( {\bf r},t\right) \equiv \E$ is the slowly varying amplitude,
$\omega=ck$ is the central frequency of probe photons, ${\bf 
k}=\widehat{{\bf z}}k$
is the wave vector and $\widehat{\bf e}\perp \widehat{\bf z}$
is the unit polarization
vector. The dimension of $\E$ is such that the operator $\E^{\dagger }\E$
represents the number density of probe photons.

The atoms (initially in the ground level $g$) are moving along the $x$ axes
with an average speed ${\bf v}_{0}{\bf =}\hbar {\bf k}_{0}/m$ and the
kinetic energy $\omega _{at}=\hbar ^{2}k_{0}^{2}/2m$. The atomic velocities
are assumed to be spread over a narrow range around $v_{0}$, so that we
can introduce slowly varying atomic amplitudes as: $\Phi _{g}=\Psi
_{g}e^{i\left( \omega _{at}t-{\bf k}_{0}\cdot{\bf r}\right) }$, $\Phi 
_{e}=\Psi
_{e}e^{i\left( \omega _{at}+\omega \right) t-i\left( {\bf k}_{0}+{\bf k}%
\right) \cdot {\bf r}}$ and
$\Phi _{q}=\Psi _{q}e^{i\left( \omega _{at}+\omega
-\omega _{c}\right) t-i\left( {\bf k}_{0}+{\bf k}-{\bf k}_{c}\right)\cdot
{\bf r}%
}$.

Consider first storing of the probe by the first control laser. The 
following equations
hold for the slowly varying electromagnetic and matter field operators:
\begin{eqnarray}
\left( \frac{\partial }{\partial t}+c\frac{\partial }{\partial z}\right) \E%
&=& ig\Phi _{g}^{\dagger } \Phi _{e},  \label{eq-rad}\\
\left(\frac{\partial}{\partial t}+v_{0}\frac{\partial }{\partial x}\right)
{\Phi}_{g}&=&i g\E^{\dagger
}\Phi _{e},
\label{eq-at-g2}
\\
\left(\frac{\partial}{\partial t}+ i\Delta+v_{0}\frac{\partial }{\partial 
x}%
+v_{r}\frac{\partial }{\partial z}\right) \Phi _{e}
&=&i\Omega _{1}\Phi _{q}+i g\E\Phi _{g},
\label{eq-at-e2}
\\
\left(\frac{\partial}{\partial t}+ i\delta+v_{0}\frac{\partial }{\partial 
x}%
+\Delta v_{r}\frac{\partial }{\partial z}\right) \Phi _{q}
&=&i\Omega _{1}\Phi _{e},
\label{eq-at-q2}
\end{eqnarray}
where $g=\mu \sqrt{ck/2\varepsilon _{0}\hbar }$ characterizes the strength
of the radiation-matter coupling, and
$\delta =\omega _{q}+\omega _{c}-\omega -\omega _{g}+\Delta \omega _{r}$
and
$\Delta =\omega _{e}-\omega -\omega _{g}+\omega _{r},$
are the two- and single-photon detunings,
$v_{r}{=}\hbar k/m$ is the atomic recoil velocity due to absorption of
a probe photon, $\hbar\omega _{r}=\hbar k^{2}/2m$ is the 
recoil frequency and $\Delta
\omega _{r}=\hbar \left( k-k_{c}\right) ^{2}/2m$. The probe and control
beams are assumed to be copropagating. In this case the overall 
recoil velocity $\Delta v_{r}{\bf =}
\hbar \left( k-k_{c}\right) /m$ is small and can be neglected in eq.(\ref
{eq-at-q2}). Dissipation of the excited state $|e\rangle$ can be
included in (\ref{eq-at-e2}) replacing $\Delta$
by $\Delta-i\gamma$ and adding the appropriate noise
operator.

Assume that the probe field $\E$ is sufficiently weak so that one can
disregard the
depletion of the gound level $|g\rangle$. 
Neglecting the last term in eq.(\ref
{eq-at-g2}), one has
\begin{equation}
\frac{\partial}{\partial t}{\Phi}_{g}({\bf r},t)
=-v_{0}\frac{\partial }{\partial x}\Phi _{g}({\bf r},t).  \label{eq-at-g2a}
\end{equation}
It is convenient to introduce the field operators describing annihilation of
atomic and spin excitations (excitons):
\begin{equation}
\psi _{u}\equiv \psi _{u}({\bf r},t)=n^{-1/2}\Phi _{g}^{\dagger }\Phi _{u}
\label{psi-x}
\end{equation}
with $u=e, q$, where $n\equiv n(z) =\langle \Phi _{g}^{\dagger }
\Phi _{g}\rangle$ describes the spatial profile of the number-density of
ground-state atoms.
Exploiting relation (\ref{eq-at-g2a}),
the equations of motion for the new operators have the form of eqs.(%
\ref{eq-rad}), (\ref{eq-at-e2}) and (\ref{eq-at-q2}) subject to the
replacement $\Phi _{e}\rightarrow \psi _{e}$, $\Phi _{q}\rightarrow
\psi _{e}$, and $\Phi _{g}\rightarrow n^{1/2}$. Note that the exciton
operators $\psi _{e}$ and $\psi _{q}$
obey approximately Bose-commutation relations
even through
the constituent atoms may be fermions.

Let us consider the case of 
exact two-photon resonance ($\delta =0$). This is well
justified as the two-photon Doppler shift due to longitudinal and
transversal velocity spreads of the atoms is strongly diminished due to
the chosen geometry.
Neglecting
terms containing $\psi _{e}$ and $\dot{\psi}_{e}$ in eq. (\ref{eq-at-e2}), one
arrives at the adiabatic approximation relating $\psi _{q}$ to the electric
field amplitude $\E(\mathbf{r},t)$ as:
\begin{equation}
\psi _{q}({\bf r},t)=-gn^{1/2}\widetilde{\E}({\bf r},t)
\label{psi-q1}
\end{equation}
with
%
%
%\begin{equation}
$ \widetilde{\E}({\bf r},t)=
\E ({\bf r},t)/\Omega _{1}(x)=-\psi _{q}({\bf r},t)/gn^{1/2}
$
%\label{E-tilde}
%\end{equation}
%
%
being an auxiliary field.
Assuming a stationary flow of atoms in the $x$-direction,
the atomic density $n\equiv n\left( z\right) $ does not depend on $x$,
and thus
eqs. (\ref{eq-at-q2}) and (\ref{psi-q1}) yield
\begin{equation}
\psi _{e}({\bf r},t)=i\frac{gn^{1/2}}{\Omega _{1}}\left( v_{0}\frac{\partial 
}{%
\partial x}+\frac{\partial }{\partial t}\right) \widetilde{\E}({\bf r},t).
\label{psi-e1}
\end{equation}
On the other hand, the Rabi frequency $\Omega _{1}\equiv \Omega _{1}\left(
x\right) $ is $z$ independent, so eqs. (\ref{eq-rad}) and (\ref{psi-e1})
lead to the following equation for $\widetilde{\E}({\bf r},t)$:
\begin{equation}
\left( \frac{\partial }{\partial t}+\frac{c}{1+n_g}\frac{\partial }{%
\partial z}+\frac{v_{0} n_g}{1+n_{g}}\frac{\partial }{\partial x}\right)
\widetilde{\E}({\bf r},t)=0,  \label{eq-rad'0}
\end{equation}
with $n_g(x,z)=g^2 n(z)/ \Omega_1^2(x)$ being the group index.
Assuming slow propagation,
the group index is much larger than unity, so that eq. (\ref{eq-rad'0})
simplifies to
\begin{equation}
\left( \frac{\partial }{\partial t}+\widetilde{v}_{g}\left( z\right) a\left(
x\right) \frac{\partial }{\partial z}+v_{0}\frac{\partial }{\partial x}%
\right) \widetilde{\E}({\bf r},t)=0,
\label{eq-rad'}
\end{equation}
where  $ v_{g} \equiv \widetilde{v}_{g}\left(
z\right) a\left( x\right) =c/n_g(x,z) $ is the group velocity ($v_{g} \ll 
c$).
The dimensionless quantity
$a\left( x\right) \equiv a_{1}\left( x\right) =\left[ \Omega _{1}\left(
x\right) /\Omega _{1}\left( x_{1}\right) \right] ^{2}  \label{a}
$ characterizes the spatial shape of the first control laser centered at
$x=x_{1}=0$.
The $z$ dependence of $\widetilde{v}_g\left(
z\right) \equiv v_{g}\left( z,x_{1}\right) $ emerges through the atomic
density $n\equiv n\left( z\right) $.
It is noteworthy that $\E({\bf r},t)=-\psi _{q}(
{\bf r},t)\sqrt{v_{g}/c}$, so the ratio between the number density
of photons and the spin excitations is given by the relative group velocity
$v_{g}/c$. In other words, the slowly
propagating probe beam is made of the EIT (dark state) polaritons
\cite{Fleisch:00,Juz:02,Fleisch:02}
comprising predominantly the spin excitations.

Assume that the spatial width of the first control beam $\Delta x_{c1}$ is
much larger than that of the incomming probe beam $\Delta x_{p}$
centered at $x=x_{1}=0$.
Under this condition, the solution of eq.(\ref{eq-rad'}) can be expressed in 
terms of the incoming electric field at an entry point $z=z_{1}$ as
\begin{equation}
\widetilde{\E}({\bf r},t)=\frac{\E\bigl[ \xi \left( x,z\right) ,z_{1},\tau
\left( t,x,z\right) \bigr] }{\Omega _{1}\left( x_{1}\right) },
\label{E-time-dep1}
\end{equation}
where
$\xi \left( x,z\right) =\int_{0}^{x}a\left( x^{\prime }\right)
dx^{\prime }-v_{0}\int_{z_{1}}^{z}\left[ \widetilde{v}_{g}
\left(z^\prime\right) \right] ^{-1}dz^\prime $,
and $\tau \left( t,x,z\right) =t+ \xi \left( x,z\right) /v_{0}-x/v_{0}$
obey the proper boundary conditions:
$ \xi \left( x,z=z_1\right) \approx x$ and
$\tau \left( t,x,z=z_1\right) \approx t$ for the incoming probe field.

Using the definition of $\widetilde \E$  and eq.(\ref{E-time-dep1}), one 
finds the temporal
and spatial behavior of the operators for\ electric field and spin
excitations:
\begin{eqnarray}
\E\left( \bf r,t\right) &=&\sqrt{|a\left( x\right)| }\, \E\left[ \xi \left(
x,z\right) ,z_{1},\tau \left( t,x,z\right) \right],  \label{E-time-dep2}\\
\psi _{q}({\bf r},t)& =& -\frac{gn^{1/2}}{\Omega _{1}\left( x_1 \right) }\E
\left[ \xi \left( x,z\right) ,z_{1},\tau \left( t,x,z\right) \right].
\label{psi-q-time-dep}
\end{eqnarray}
Equations (\ref{E-time-dep2}) and (\ref{psi-q-time-dep}) define the
electric and spin components of the EIT polariton.
A set of trajectories for such a polariton in the $x-z$ plane
is given by $\xi \left( x,z\right)=\xi_0$,
with $\xi_0=0$ corresponding to the central trajectory.
Due to the motion of the atoms parallel to the  $x$ axis, the polariton is
dragged into that direction. Following the spatial profile $a(x)$ of the
coupling beam, its velocity component in the $z$ direction, $\widetilde v_g(z)
a(x)$, is further reduced. As soon as $\widetilde v_g(z) a(x)$ becomes less than $v_0$
the flow  of excitation is predominatly determined by the velocity
of the atoms $v_0$ in the $x$ direction.
If the atomic beam is optically thick in the $z$direction
($\xi \left( \infty,z_{max}\right)<0$), the propagation direction of the
polariton may be completely converted from $\hat{\bf z}$ to $\hat{\bf x}$.
The probe field is then stored in the form of a spin excitation that moves with
the atoms parallel to the $x$-axis and is centered at $z=z_{\infty}$, which is a
solution of $\xi \left(x\to \infty, z_{\infty}\right) =0$.
This is illustrated in figs.2 and 3 where we have shown the amplitude of the
electric field at different instances of time. Also shown is the
effect of the second regeneration laser, which will be discussed later on.

%%%%%%%%%%%%%%%%%%%%%%%%%%%%%%%%%%%%%%%%%%%%%%%%%%%%%%%%%%%%%%%%%%%%%%%%%
%
%
\begin{figure}[ht]
\begin{center}
\includegraphics[width=7cm]{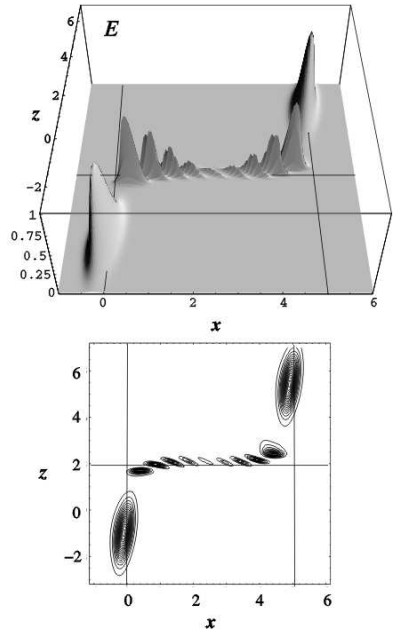}
\caption{Propagation of a probe pulse in a moving EIT medium
at times $t=0, 5, 10, 15, 20, 25, 30, 35, 40, 45, 50$
illuminated by a pair of control lasers propagating in the $+z$
direction at $x_1=0$ and $x_2=5$. The control lasers have equal amplitudes
and Gaussian profiles with unity width. The flow and group velocities are 
$v_0=0.1$, and
$\widetilde v_g(z)=(1-0.95 \exp\{-(z-2)^2\})$.}
\label{stream1}
\end{center}
\end{figure}
%
%%%%%%%%%%%%%%%%%%%%%%%%%%%%%%%%%%%%%%%%%%%%%%%%%%%%%%%%%%%%%%%%%%%%%%%%%%%

%%%%%%%%%%%%%%%%%%%%%%%%%%%%%%%%%%%%%%%%%%%%%%%%%%%%%%%%%%%%%%%%%%%%%%%%%
%
%
\begin{figure}[ht]
\begin{center}
\includegraphics[width=7cm]{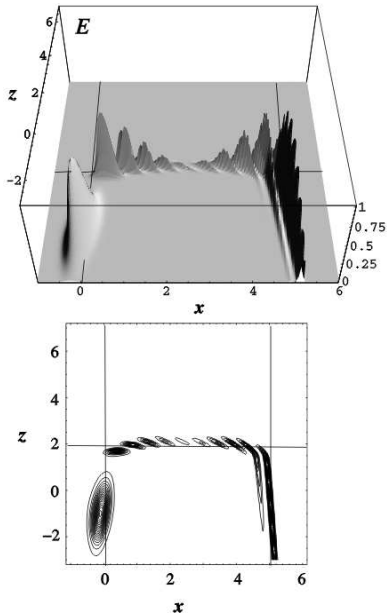}
\caption{Propagation of a probe pulse in a moving EIT medium
for conditions of fig.\ref{stream1} but with a counterpropagating
(i.e. in $-z$ direction) second control field.}
\label{stream2}
\end{center}
\end{figure}
%
%%%%%%%%%%%%%%%%%%%%%%%%%%%%%%%%%%%%%%%%%%%%%%%%%%%%%%%%%%%%%%%%%%%%%%%%%%%

As one can see from the figures, the dragging of the polariton along with the
moving atoms leads to a deformation of the excitation. For large values of $x$,
such that $a(x)\to 0$, one has in the vicinity of $z=z_{\infty}$
\begin{eqnarray}
\xi \left(x\to \infty, z\right)
\approx - \left(z-z_{\infty}\right) 
\frac{v_{0}}{\widetilde{v}_{g}(z_{\infty})},
\label{fi-limit-approx}\\
\tau \left(t,x\to \infty, z\right)
\approx - 
\frac{z-z_{\infty}}{\widetilde{v}_{g}(z_{\infty})}-\frac{x-x_0}{v_0},
\end{eqnarray}
where $x_0 \equiv x_0 \left( t \right) =v_0 t$ is the "center of mass" of a
stored polariton moving in the $x$ direction.  Thus, if $\Delta x_{p}$ and
$2\Delta \tau _{p} $ denote the half-width (in $x$ direction) and the duration
of the input probe pulse, the dimensions of the spin wave are in the limit
$x\to\infty:$
\begin{eqnarray}
\Delta x_s \approx  v_0 \Delta \tau _{p},\qquad
\Delta z_s \approx \Delta x_p
\frac{\widetilde{v}_{g}(z_{\infty})}{v_{0}},\label{Delta-z}
\end{eqnarray}
where we have assumed that $v_{0}\ll \Delta x_{p}/\Delta \tau_{p}$.  Since the
spatial profile of the polariton in the $y$ direction is not changed, the
transfer from a electromagnetic to a matter-wave pulse is associated with a
spatial compression of the excitation volume by a factor
$\widetilde{v}_{g}(z_{\infty})/{c}$.  From eq.(\ref{Delta-z}) one can also
easily obtain a necessary condition for a complete transfer of the
electromagnetic excitation to the atomic beam: Since $\Delta z_s$ should be less
than the half width of atomic beam $\Delta z_{\rm atom}$, one finds a minimum ratio of
$v_0$ to the group velocity at the center
\begin{equation}
\frac{v_0}{\widetilde{v}_{g}(z_{\infty})}
\gg \frac{\Delta x_{p}}{\Delta z_{\rm atom}}.
\end{equation}
If this condition is fulfilled a pulse is completely stored within the atom beam
and does not exit on the back side.

Consider now the regeneration of the probe beam by a spatially separated, i.e.
nonoverlapping, second control laser centered at $x=x_{2}$ and characterized by
a Rabi frequency $\Omega _{2}\left( x\right) $.  In this case, the previous
equations (\ref{eq-rad'}) - (\ref{fi-limit-approx}) describe the propagation of
a probe beam within the entire system, subject to the following replacement:
$a\left( x\right) \longrightarrow a\left( x\right) =a_{1}\left( x\right) 
\pm a_{2}\left( x\right) $
where $a_{2}\left( x\right) =\left[ \Omega _{2}\left( x\right) /\Omega
_{2}\left( x_{1}\right) \right] ^{2}$ characterizes the shape of the second
control laser. The upper (lower) sign corresponds to the case where the second
control laser propagates in the same (opposite) direction, as compared to the
first one \cite{footn:2}. This is because the radiative group velocity of the
regenerated probe beam changes sign in the latter case. It is noteworthy that
the reversed probe beam experiences a slight shift out of the EIT resonance
\cite{Juz:02}. Yet such a shift can be neglected in the case where the initial
control and probe beams are copropagating. Note also that the previously
considered regeneration of a probe beam is due to the temporal switching of the
control laser
\cite{Fleisch:00,Hau:01,Lukin:01,Scully:01,Juz:02,Fleisch:02,Scully:02}, whereas
now the regeneration is induced by a second spatially separated control laser
that can be stationary.

Finally, let us analyze the output field at a spatial point where $z=z_{2}$.  The
maximum electric and spin fields are then concentrated at $x=x_{2}$, for which
$\xi \left( x_2,z_2\right) =0$. In the vicinity of this point, one has
$\xi \left( x,z_{2}\right) \approx \pm a_{2}\left( x_{2}\right) 
\left( x-x_{2}\right)$ i.e. the output field has a width $\Delta x_{p2}= \Delta
x_{p1}/  a_{2}\left(x_{2}\right)$.

In summary, we have investigated a different scheme of storing and releasing a beam
of probe light in a moving atomic medium illuminated by two spatially separated
control lasers depicted in fig.\ref{setup}. Beyond the area illuminated by the
first control laser, the probe beam transforms into a beam of pure spin
excitations moving along the $x$ axis, as one can see from
figs.\ref{stream1},\ref{stream2}.  The regeneration of the probe beam is
accomplished by applying the second continuous control laser.  Depending on the
direction of the latter, the restored probe beam moves either parallel
(fig.\ref{stream1}), or antiparallel (fig.\ref{stream2}) to the initial probe
beam.

In contrast to the previous schemes
\cite{Fleisch:00,Hau:01,Lukin:01,Scully:01,Juz:02,Fleisch:02,Scully:02},
storing and releasing of a probe beam can be now accoplished without switching
"off" and "on" of a control laser.  This is advantageous if one does not know the
exact time of arrival of the probe photons, e.g. if the latter are created via
spontaneous processes.


\begin{thebibliography}{99}
\bibitem{footn:2}  The second control laser should be properly polarized in 
order to drive the same transition $q\rightarrow e$.
\end{thebibliography}
\end{document}